\documentstyle[pra,aps,twocolumn]{revtex}
\begin{document}
\draft
\title{Fidelity for multimode thermal squeezed states}

\author{Gh.-S. Paraoanu\thanks{On leave from the Department of Theoretical Physics, Institute of Atomic Physics, PO Box MG-6, R-76900, Bucharest--Magurele, Romania.}}

\address{Department of Physics, University of Illinois at Urbana--Champaign, 1110 W. Green St., Urbana, IL 61801, USA; e-mail: paraoanu@physics.uiuc.edu}

\author{Horia Scutaru }

\address{Department of Theoretical Physics, Institute of
Atomic Physics, PO BOX MG-6,
 R-76900 Bucharest-Magurele, Romania; e-mail: scutaru@theor1.theory.nipne.ro}

\maketitle
\begin{abstract}

In the theory of quantum transmission of information the concept of fidelity 
plays a fundamental role. An important class of channels, which can be 
experimentally realized in quantum optics, is that of Gaussian quantum 
channels. In this work we present a general formula for fidelity in the 
case of two arbitrary Gaussian states. From this formula one can get a 
previous result \cite{scut1},
for the case of a single mode; or, one can apply it to obtain a closed
compact expression for multimode thermal states. 
The concept of fidelity used in this paper is
the standard one \cite{bu1,u1,j1}. It can be defined by
\[
F\left(\rho_1, \rho_2\right) \stackrel{{\mathrm def}}{=} 
\max_{|\psi_1>, |\psi_2>}\left|\left<\psi_1|\psi_2\right>\right|^2,
\]
where $|\psi_i>$, $i = 1,2$ are purifications of the density matrices
$\rho_{i}$.

\end{abstract}

\pacs{03.65.Bz, 42.50.Dv, 89.70.+c}
\narrowtext
\newtheorem{definition}{Definition}
\newtheorem{theorem}{Theorem}
\newtheorem{lemma}{Lemma}
\newtheorem{conclusion}{Corollary}
\section{Introduction}

Within recent years, the quantum theory of information, an extension of 
the classical theory of information to the quantum realm, has emerged as 
a fascinating research field. A great deal of effort has been devoted to 
the issue of transmitting a state through noisy quantum channels despite 
the quantum-mechanical 
uncertainties in our knowledge about that state. This is a different problem 
from the classical situation, where the states are mutually exclusive and the 
input system may remain in its initial state; 
in the quantum case the states are density operators on a 
Hilbert space and the non-cloning theorem \cite{clo} precludes 
the input system, in general, to retain its original state. 
Moreover, for a noisy quantum channel the state is subject to a 
decoherence process, due to the interaction with an external environment, 
which further decreases the reliability of information processing. 
Thus, a fundamental problem is to extend the classical encoding and 
decoding procedures to quantum channels and to define an upper limit 
(the channel capacity) to the amount of quantum information that can be 
transmitted 
with an arbitrary high fidelity. Recently, work \cite{gau} has been focused on 
quantum channels which use Gaussian  
(or quasi-free, \cite{hol1,scut2}) states 
for the transmission of 
information. In this case, a first issue to be raised is how 
to  calculate
the fidelity given two mixed Gaussian states; partial answers, 
depending on the particular type of mixtures under consideration 
have already been given in the literature \cite{scut1,tw,pa}. 
In this paper, we will give a general formula
for the fidelity of two quasi-free states, and show that the previous results
can be obtained as particular cases. 
The formula can also be applied to some interesting cases, 
such as two-mode systems \cite{prep}.

Let $\rho_1$ and $\rho_2$ two density operators which describe 
two mixed states. The transition probability $P(\rho_1,\rho_2)$ 
has to satisfy the following natural axioms:

\begin{enumerate}
\item $P(\rho_1,\rho_2) \leq 1$ and $P(\rho_1,\rho_2) = 1$ 
if and only if $\rho_1 = \rho_2$;

\item  $P(\rho_1,\rho_2) = P(\rho_2,\rho_1)$;

\item  If $\rho_1$ is a pure state, $\rho_1 = |\psi_1><\psi_1|$ then 
\newline $P(\rho_1, \rho_2) = <\psi_1|\rho_2|\psi_1>$;

\item  $P(\rho_1, \rho_2)$ is invariant under unitary transformations on the 
state space;

\item  $P\left(\rho_1|_{\cal A},
 \rho_2|_{\cal A}
\right) \geq P(\rho_1, \rho_2)$ for any complete subalgebra 
of observables ${\cal A}$;

\item $P(\rho_1\otimes\sigma_1, \rho_2\otimes\sigma_2 ) = 
P(\rho_1, \rho_2)P(\sigma_1, \sigma_2)$.
\end{enumerate}

Uhlmann's transition probability for mixed states \cite{u1}
\begin{equation}
P(\rho_1, \rho_2) = 
\left[Tr\left(\sqrt{\rho_1}\rho_2\sqrt{\rho_1}\right)^{1/2}\right]^2\label{2}
\end{equation}
satisfies properties 1--6. The fidelity is defined by 
$F(\rho_1, \rho_2) = P(\rho_1, \rho_2)$. A detailed analysis of the 
structure of the transition probability was hampered by the factors 
containing square roots in Eq. (\ref{2}).
Due to technical difficulties in the computation of fidelities, 
few concrete examples of analytic calculations are known. Until recently, 
all the results were obtained only for finite-dimensional Hilbert spaces 
\cite{hu,hu1,dit}. The first results in an infinite-dimensional Hilbert 
space were recently obtained by
Twamley \cite{tw} for the fidelity of two thermal squeezed states and by 
Paraoanu and Scutaru \cite{pa} for the case of two displaced thermal states. 
In \cite{scut1} Scutaru has developed another calculation method which allowed 
getting the result for the case of two displaced thermal squeezed states
in a coordinate-independent form.

Let $(E, \sigma)$ be a phase space i.e. a vector space with
a symplectic structure $\sigma$. Then the commutation relations  
on $(E, \sigma)$ acting in a Hilbert space ${\cal H}$ are defined 
by a continuous family of unitary operators $\{ V(u), u \in E \}$
on ${\cal H}$ which satisfy the Weyl relations \cite{hol1,scut2}:
\begin{equation}
V(u)V(v)=\exp\left\{{i \over 2} \sigma(u,v)\right\}V(u+v).
\end{equation}
Hence the family 
$\{V(tu), - \infty < t < \infty \}$ for a fixed
$u \in E$
is a group of unitary operators.

Then by the Stone theorem 
\begin {equation}
V(u) = \exp\{i R(u)\},
\end{equation}
where $R(u)$ is a selfadjoint operator.
From the Weyl relations we have
\begin{eqnarray}
&&
\nonumber
\exp\{itR(u)\}\exp\{isR(v)\}=\\
&&
\nonumber
\exp\{its\sigma(u,v)\}\exp\{isR(v)\}\exp\{itR(u)\}.\\
&&
\nonumber
\end{eqnarray}
By differentiation and taking $t=s=0$ one obtains
\begin{equation}
[R(u),R(v)]= -i\sigma(u,v)I.
\end{equation}
The operators $\{R(u), u \in E\}$ are called 
cannonical observables.

The phase space $E$ is of even real dimension $2n$ 
and there exist in $E$
symplectic bases of vectors $\{e_j, f_j\}_{j=1,...,n}$, i.e.
reference systems such that $\sigma(e_j,e_k) = \sigma(f_j,f_k)=0$
and $\sigma(e_j,f_k) = - \sigma(f_k,e_j) =
\delta_{jk}$, $j,k=1,...,n$.
The coordinates $(\xi^j,\eta^j)$ of a vector $u\in E$ in a
symplectic basis, $ u= \sum_{j=1}^n(\xi^je_{j}+\eta^jf_{j}) $, are
called symplectic coordinates. The well known coordinate and momentum
operators are defined by $Q_{k}=R(f_{k})$ and $P_{k}=R(e_{k})$ for
$k=1,2,...,n$. Then the canonical observables $R(u)$ are linear
combinations of the above defined coordinate and momentum operators:
$R(u)= \sum_{j=1}^n(\xi^jP_{j}+\eta^jQ_{j}) $.
 
There is a one-to-one
correspondence between the symplectic bases and the linear
operators $J$ on $E$ defined by $Je_k=-f_k $ and $Jf_k=e_k$,
$k=1,...,n$. The essential properties of these operators are:
$\sigma(Ju,u)\geq 0$, $ \sigma(Ju,v)+\sigma(u,Jv)=0$ ($u,v \in E$ and
$J^2=-I$, where $I$ denotes the identity operator on $E$). 
Such operators
are called complex structures.
In the following
we shall use the matriceal notations with $u\in E$ as column vectors.
Then $\sigma(u,v)=u^TJv$ and the scalar product is given by 
$\sigma(Ju,v)=
u^Tv, u,v\in E$. A linear operator $S$ on $E$ is called a
symplectic operator if $S^TJS=J$. When $S$ is a symplectic operator
then $S^T$ and $S^{-1}$ are also symplectic operators. The
group of all symplectic operators $Sp(E,\sigma)$ is called the
symplectic group of $(E,\sigma)$. The Lie algebra of $Sp(E,\sigma)$
is denoted by $sp(E,\sigma)$ and its elements are operators $R$
on $E$ with the property: $(JR)^T=JR$. Hence an operator $R$ on $E$
belongs to $Sp(E,\sigma) \cap sp(E,\sigma)$ iff $R^2=-I$.
If $J$ and $K$ are two complex
structures, there exists a symplectic transformation $S$
such that $J = S^{-1}KS$. 
For any symplectic operator $S$ we can define a new system
of Weyl operators $\{V(Su); u \in E\}$. Then from a well known
result on the unicity of the the systems of Weyl operator up to
a unitary equivalence it follows that there exists a unitary
operator $U(S)$ on ${\cal H}$ such that $V(Su)=U(S)^{\dag}V(u)U(S)$.

For any nuclear operator $O$ on ${\cal H}$ one defines the
characteristic function

\begin{equation}
CF_{u}(O)= Tr O V(u), ~~~~~u \in E.
\end{equation}
We give the properties of the characteristic function
which are important in the following \cite{hol1}:

\begin{enumerate}
\item $CF_{0}(O)=TrO$;
\item $CF_{u}\left[V(v)^{*}OV(v)\right]=
CF_{u}\left[O\exp{i\sigma(v,u)}\right]$;
\item $CF_{u}(O_{1}O_{2})=
\newline {1 \over (2 \pi)^{n}} \int CF_{v}(O_{1})
CF_{u-v}(O_{2})\exp{{i \over 2} \sigma(v,u)} dv$;
\item $CF_{Su}(O)=CF_{u}(U(S)OU(S)^{\dag})$.
\end{enumerate}

\section{Multimode thermal squeezed states}
The multimode thermal squeezed states are defined
by the density operators $\rho$ whose characteristic functions
are Gaussians \cite{scut1,hol1,scut2}
\begin{equation}
CF_{u}(\rho)= \exp\left\{-{1 \over 4}u^{T}Au\right\}.
\end{equation}
where $A$ is a $2n \times 2n$ positive definite matrix,
called correlation matrix. From the last property of the
characterisic function, enumerated above, it follows that:
\begin{equation}
A_{U(S) \rho U(S)^{\dag}}=S^TA_{\rho}S
\end{equation}

Because the correlation matrix A is positive definite it
follows \cite{scut2,fol} 
that there exists $S \in Sp(E,\sigma)$ such that
\begin{equation} 
A = S^{T}{\cal D}S 
\end{equation}
where
${\cal D}=
\left(\matrix{D&0\cr 0&D\cr}\right)$
and $D \geq I$ is a diagonal $n\times n$ matrix.
   The most general real symplectic transformation $S\in Sp(E,\sigma)$
has \cite{scut2,bal}
the following structure:
\begin{equation}
S = O{\cal M}O^{'} 
\end{equation}
where
\begin{equation}
{\cal M}=\left(\matrix{M&0\cr 0&M^{-1}\cr}\right)
\end{equation}
and $O$, $O^{'}$ are symplectic and orthogonal $(O^{T}O=I)$
operators, and where $M$ is a diagonal $n\times n$ matrix.
Various particular kinds of such matrices are obtained taking $O$,
$O^{'}$, ${\cal D}$ or ${\cal M}$ to be equal or proportional to 
the corresponding 
identity operator.
A pure squeezed state is obtained when ${\cal D}=I$. If this condition
is not satisfied, the state is a mixed state called thermal squeezed
state \cite{eza}. When ${\cal M}=I$ there is no squeezing and the 
correspondig states
are pure coherent states or thermal coherent states. All these states
have correlations between the different modes produced by the
orthogonal symplectic operators $O$ and $O^{'}$.
As a consequence the most general form of a correlation matrix $A$ 
is given by:  
\begin{eqnarray}
A = O^{'T}{\cal M}O^{T}{\cal D} O{\cal M}O^{'} 
\end{eqnarray}
From the property 3 of the characteristic function
we have for two density operators $\rho_{1}$ an
$\rho_{2}$
\begin{eqnarray}
&&
\nonumber
CF_{u}(\rho_{1}\rho_{2})=
\left[det\left({A_{1}+A_{2} \over 2}\right)^{-{1 \over 2}}\right]\\
&&
\nonumber
\exp{\left\{-{1 \over 4}u^{T}\left[A_{2}-
(A_{2}-iJ)(A_{1}+A_{2})^{-1}(A_{2}+iJ)\right]u\right\}}.
\end{eqnarray}
When $\rho_{1}=\rho_{2}$ we have
\begin{equation}
CF_{u}(\rho^2)= (detA)^{-{1 \over 2}}\exp
\left\{-{1 \over 4}u^T\left({A-JA^{-1}J \over 2}\right)u\right\}.
\end{equation}
A state $\rho$ is pure iff $\rho^2=\rho$.
Then from the equality 
$CF_{u}(\rho^2)=CF_{u}(\rho)$ 
it follows that a Gaussian
state is pure iff
\begin{equation}
A=-JA^{-1}J,
\end{equation}
i.e. a Gaussian state is pure iff $JA \in Sp(E,\sigma)$.
Analogously, for a mixed state $\rho^2 < \rho$.
Then $CF_{u}(\rho^2) < CF_{u}(\rho)$ and as 
a consequence ${A-JA^{-1}J \over 2} > A$. 
Hence for any Gaussian state the correlation
matrix $A$ must satisfy the folloving restriction
\begin{equation}
A \leq -JA^{-1}J.
\end{equation}

\section{The characteristic function of the square root of a density matrix}

Let us suppose that the characteristic function of the
Hilbert-Schmidt operator $\sqrt{\rho}$ of a Gaussian state is, 
up to a numerical factor, also
a Gaussian function with the correlation matrix $\Phi(A)$,
\begin{equation}
CF_{u}(\sqrt{\rho})= K \exp\left\{-{1 \over 4}u^T\Phi(A)u\right\}.
\end{equation}
Then from the equality $\rho = \sqrt{\rho}\sqrt{\rho}$
we obtain
\begin{eqnarray}
&&
\nonumber
K^2\left(det(\Phi(A))\right)^{-{1 \over 2}}\\
&&
\nonumber
\exp\left\{-{1 \over 4}
\left({\Phi(A)-J\Phi(A)^{-1}J \over 2}\right)u\right\}= \\
&&
\nonumber
\exp\left\{-{1 \over 4}u^TAu\right\}.\\
\end{eqnarray}
Hence
\begin{equation}
K^2=\sqrt{det\Phi(A)},
\end{equation}
and
\begin{equation}
\Phi(A)-J\Phi(A)^{-1}J=2A.
\end{equation}
The last equation has the solution
\begin{equation}
\Phi(A)=A(I+\sqrt{I+(JA)^{-2}}).
\end{equation}
This is a new proof of a result obtained in \cite{hol1}. The
advantage of this proof is given by the fact that it does not
require the choice of a special basis in $E$. If we take a 
symplectic basis in {E}
such that $J=\left(\matrix{0&I \cr -I&0 \cr}\right)$ then 
 $J{\cal D}
={\cal D}J$ and
from this equation and from the equations 
$A = S^T{\cal D} S$, $JS^T = S^{-1}J$ it follows that 
$(JA)^{-2} = S^{-1}{\cal D}^{-2}S$.
Hence $I+(JA)^{-2} =S^{-1}(I-{\cal D}^{-2})S$ and
\begin{eqnarray}
A\sqrt{I+(JA)^{-2}}=
S^{T}(\sqrt{{\cal D}^{2}-I})S
\end{eqnarray}

\section{The general formula for the fidelity of Gaussian states}

The fidelity $F(\rho_{1},\rho_{2})$ for two density operators
$\rho_{1}$ and $\rho_{2}$ is defined by
\begin{equation}
F(\rho_{1},\rho_{2})=
Tr\left(\sqrt{\sqrt{\rho_{1}}\rho_{2}\sqrt{\rho_{1}}}\right).
\end{equation}
As we have seen in section II the characteristic function of a 
product of operators whose characteristic functions are Gaussians
is also a Gaussian. In section III we have obtained a simple formula for 
the characteristic function of the square root of a density operator
whose characteristic function is a Gaussian. Hence we can find a simple 
formula for the characteristic function of the operator $\sqrt{\rho_{1}}
\rho_{2}\sqrt{\rho_{1}}$:  
\begin{equation}
CF_{z}(\sqrt{\rho_{1}}\rho_{2}\sqrt{\rho_{1}})=\sqrt{L}\exp\left\{-{1 \over 4}
z^T{\cal O}z\right\},
\end{equation}
where
\begin{eqnarray}
&&
\nonumber
L^{-1}=det\Phi(A_{1})^{-1} 
det\left({\Phi(A_{1})+A_{2} \over 2}\right)\\
&&
\nonumber
det\left({A_{2}+\Phi(A_{1})-{\cal U}
\over 2}\right)\\
\end{eqnarray}
where ${\cal U}=(A_{2}-iJ)(\Phi(A_{1})+A_{2})^{-1}(A_{2}+iJ)$,
and
\begin{eqnarray}
&&  
\nonumber
{\cal O}= \Phi(A_{1})-(\Phi(A_{1})-iJ)[A_{2}+\Phi(A_{1})-\\
&&
\nonumber
(A_{2}-iJ)(\Phi(A_{1})+A_{2})^{-1}(A_{2}+iJ)]^{-1}(\Phi(A_{1})+iJ).
\end{eqnarray}
Then applying the result of the preceeding section
we can obtain the characteristic function of
$\sqrt{\sqrt{\rho_{1}}\rho_{2}\sqrt{\rho_{1}}}$,
\begin{eqnarray}
&&
\nonumber
CF_{z}\left(\sqrt{\sqrt{\rho_{1}}\rho_{2}\sqrt{\rho_{1}}}\right)=\\
&&
\nonumber
\left[Ldet\Phi({\cal O})\right]^{{1 \over 4}}\exp\left\{-{1 \over 4}z^T\Phi
({\cal O})z\right\}.\\
\end{eqnarray}
From this formula and the property 1 of the characteristic function
we obtain
\begin{equation}
F(\rho_{1},\rho_{2})=\sqrt{Ldet\Phi({\cal O})}. 
\end{equation}
We remark that 
\begin{equation}
det\Phi({\cal O})=det{\cal O}det\left[I+\sqrt{I+(J{\cal O})^{-2}}\right].
\end{equation}
In order to simplify the formula for fidelity we observe that
\begin{eqnarray}
&&
\nonumber
t_{ijk}=Tr\rho_{i}\rho_{j}\rho_{k}=
det\left({A_{i}+A_{j} \over 2}\right)\\
&&
\nonumber
det\left[{A_{j}+A_{k}-(A_{j}-iJ)
(A_{i}+A_{j})^{-1}(A_{j}+iJ) \over 2}\right],
\end{eqnarray}
and that $t_{123}=t_{231}=t_{312}$.
If we take in this last identity $\Phi(A_{1})$
instead of $A_{1}$ then we obtain
\begin{eqnarray}
&&
\nonumber
det\left[{\Phi(A_{1})+A_{2} \over 2}\right]\\
&&
\nonumber
det\left[{A_{2}+\Phi(A_{1})-{\cal U} \over 2}\right]\\
&&
\nonumber
=det\left({A_{1}+A_{2} \over 2}\right)det\Phi(A_{1}).
\end{eqnarray}
Hence we get
\begin{equation}
L=\left[det\left({A_{1}+A_{2} \over 2}\right)\right]^{-1}.
\end{equation}

It is not evident from this general formula that the properties
1-6 of the fidelity are valid.
Let us consider the most simple one, namely the property
$F(\rho,\rho)=1$. In this case it is necessary to prove
that $\Phi({\cal O})=A$. We can choose the
complex structure $J$ to commute with the correlation
matrix $A$: $JA=AJ$. Then all operations in the formula
which gives ${\cal O}$ as a function of $A$ and $J$ can be performed
and the result is: ${\cal O}={A+A^{-1} \over 2}$ and 
$\Phi({\cal O})=A$. The next property which we shall discuss   
is the property 3 which in the case of Gaussian states becomes
\cite{scut1}: $F(\rho_{1},\rho_{2})=({A_{1}+A_{2} \over 2})^{-{1
\over 2}}$. We shall prove that ${\cal O}=A_{1}$ when $\rho_{1}$
is a pure state. First we remark that $\Phi(A_{1})=A_{1}$ and
that there is a symplectic transformation such that $A_{1}=S^TS$.
Then ${\cal O}=S^T\left\{I-2P_{-}\left[2I-4P_{+}{\cal X}^{-1}P_{-}\right]
2P_{+}\right\}S$ where
\newline $P_{+}={I+iJ \over 2}$, $P_{-}={I-iJ
\over 2}$ and ${\cal X}= (S^T)^{-1}A_{2}S^{-1}+I$. Evidently $P_{+}$
and $P_{-}$ are an orthogonal decomposition of the unit operator:
$P_{+}^2=P_{+}$, $P_{-}^2=P_{-}$, $P_{+}P_{-}=P_{-}P_{+}=0$ and
$P_{+}+P_{-}=I$. As a consequence of the orthogonality we obtain
${\cal O}=S^TS=A_{1}$.
Then $\Phi({\cal O})=\Phi(A_{1})=A_{1}=S^TS$ and $det(\Phi({\cal
O}))=det(S^T)det(S)=1$.

\section{The one mode case}

In \cite{scut1} we have obtained an expression  for the fidelity
in the one mode case. This formula can be reobtained as a consequence of
the above general formula. In the one mode case all matrices
are $2 \times 2$ matrices. For a $2 \times 2$ matrix ${\cal O}$
we have
\begin{equation}
\Phi({\cal O})= \epsilon {\cal O},
\end{equation}
where $\epsilon=1+\sqrt{1-{1 \over det{\cal O}}}$ and
$det\Phi({\cal O})=(\sqrt{det{\cal O}}+\sqrt{\det{\cal O}-1})^2$.
From these considerations it follows that
\begin{equation}
F(\rho_{1},\rho_{2})={2 \over \sqrt{det(A_{1}+A_{2})}
(\sqrt{det{\cal O}}-\sqrt{det{\cal O}-1})}.
\end{equation}
Thus  it is sufficient to compute $det{\cal O}$. We shall
denote by ${\cal P}$ the product $(detA_{1}-1)(detA_{2}-1)$.
After simple but long computations we obtain
\begin{equation}
det{\cal O}= 1 + {{\cal P} \over det(A_{1}+A_{2})},
\end{equation}
which gives the result of \cite{scut1} 
\begin{equation}
F(\rho_{1},\rho_{2})=
{2 \over \sqrt{det(A_{1}+A_{2})+
{\cal P} }-\sqrt{{\cal P}}}.
\end{equation}

\section{Multimode thermal states case} 
In the case of two thermal states with correlation
matrices $A_{i}={\cal D}_{i}$ with $i=1,2$ we have
$A_{i}J=JA_{i}$, $(i=1,2)$ and $A_{1}A_{2}=A_{2}A_{1}$.
Then $\Phi(A_{i})=A_{i}+\sqrt{A_{i}^2-I}$, $(i=1,2)$. 
Hence 
\begin{equation}
{\cal O}=(A_{1}+A_{2})^{-1}(A_{1}A_{2}+I)
\end{equation}
and
\begin{equation}
\Phi({\cal O})={(A_{1}+A_{2}) \over (A_{1}A_{2}+I)-\sqrt{(A_{1}^2-I)
(A_{2}^2-I)}}
\end{equation}
Finally
\begin{equation}
F(\rho_{1},\rho_{2})=\sqrt{det\left({2 \over (A_{1}A_{2}+I)-\sqrt{(A_{1}^2-I)
(A_{2}^2-I)}}\right)}
\end{equation}
which is the product of the fidelities of the corresponding one-mode
thermal states \cite{scut1,hol1,scut2}

\section{Conclusions}

In this paper we have provided a general formula for the calculation
of the fidelity of two Gaussian states. It is shown that, 
in the particular case 
of a single mode,
this formula reproduces the results already known in the literature,
and in the case of multimode thermal states it yields a compact
expression with a direct interpretation.

\acknowledgements

The second author acknowledges a partial financial support 
from the Grant Agency of the Romanian Academy.

\end{document}